

\hsize=6.0truein
\vsize=8.5truein
\voffset=0.25truein
\hoffset=0.1875truein
\tolerance=1000
\hyphenpenalty=500
\def\monthintext{\ifcase\month\or January\or February\or
   March\or April\or May\or June\or July\or August\or
   September\or October\or November\or December\fi}


\font\tenrm=cmr10 scaled \magstep1   \font\tenbf=cmbx10 scaled \magstep1
\font\sevenrm=cmr7 scaled \magstep1  
\font\fiverm=cmr5 scaled \magstep1   

\font\teni=cmmi10 scaled \magstep1   \font\tensy=cmsy10 scaled \magstep1
\font\seveni=cmmi7 scaled \magstep1  \font\sevensy=cmsy7 scaled \magstep1
\font\fivei=cmmi5 scaled \magstep1   \font\fivesy=cmsy5 scaled \magstep1

\font\tentt=cmtt10 scaled \magstep1
\font\tenit=cmti10 scaled \magstep1
\font\tensl=cmsl10 scaled \magstep1

\def\twelvepoint{\def\rm{\fam0\tenrm}
   \textfont0=\tenrm \scriptfont0=\sevenrm \scriptscriptfont0=\fiverm
   \textfont1=\teni  \scriptfont1=\seveni  \scriptscriptfont1=\fivei
   \textfont2=\tensy \scriptfont2=\sevensy \scriptscriptfont2=\fivesy
   \textfont\itfam=\tenit \def\it{\fam\itfam\tenit}
   \textfont\ttfam=\tentt \def\tt{\fam\ttfam\tentt}
   \textfont\bffam=\tenbf \def\bf{\fam\bffam\tenbf}
   \textfont\slfam=\tensl \def\sl{\fam\slfam\tensl} \rm
   \hfuzz=1pt\vfuzz=1pt
   \setbox\strutbox=\hbox{\vrule height 10.2pt depth 4.2pt width 0pt}
   \parindent=24pt\parskip=1.2pt plus 1.2pt
   \topskip=12pt\maxdepth=4.8pt\jot=3.6pt
   \normalbaselineskip=14.4pt\normallineskip=1.2pt
   \normallineskiplimit=0pt\normalbaselines
   \abovedisplayskip=13pt plus 3.6pt minus 5.8pt
   \belowdisplayskip=13pt plus 3.6pt minus 5.8pt
   \abovedisplayshortskip=-1.4pt plus 3.6pt
   \belowdisplayshortskip=13pt plus 3.6pt minus 3.6pt
   \topskip=12pt \splittopskip=12pt
   \scriptspace=0.6pt\nulldelimiterspace=1.44pt\delimitershortfall=6pt
   \thinmuskip=3.6mu\medmuskip=3.6mu plus 1.2mu minus 1.2mu
   \thickmuskip=4mu plus 2mu minus 1mu
   \smallskipamount=3.6pt plus 1.2pt minus 1.2pt
   \medskipamount=7.2pt plus 2.4pt minus 2.4pt
   \bigskipamount=14.4pt plus 4.8pt minus 4.8pt}

\twelvepoint



\font\titlerm=cmr10 scaled \magstep3
\font\titlerms=cmr10 scaled \magstep1 
\font\titlei=cmmi10 scaled \magstep3  
\font\titleis=cmmi10 scaled \magstep1 
\font\titlesy=cmsy10 scaled \magstep3   
\font\titlesys=cmsy10 scaled \magstep1  
\font\titleit=cmti10 scaled \magstep3   
\skewchar\titlei='177 \skewchar\titleis='177 
\skewchar\titlesy='60 \skewchar\titlesys='60 

\def\titlefont{\def\rm{\fam0\titlerm}
   \textfont0=\titlerm \scriptfont0=\titlerms 
   \textfont1=\titlei  \scriptfont1=\titleis  
   \textfont2=\titlesy \scriptfont2=\titlesys 
   \textfont\itfam=\titleit \def\it{\fam\itfam\titleit} \rm}


\def\preprint#1{\baselineskip=19pt plus 0.2pt minus 0.2pt \pageno=0
   \begingroup
   \nopagenumbers\parindent=0pt\baselineskip=14.4pt\rightline{#1}}
\def\title#1{
   \vskip 0.9in plus 0.45in
   \centerline{\titlefont #1}}
\def\secondtitle#1{}
\def\author#1#2#3{\vskip 0.9in plus 0.45in
   \centerline{{\bf #1}\myfoot{#2}{#3}}\vskip 0.12in plus 0.02in}
\def\secondauthor#1#2#3{}
\def\addressline#1{\centerline{#1}}
\def\abstract{\vskip 0.7in plus 0.35in
     \centerline{\bf Abstract}
     \smallskip}
\def\finishtitlepage#1{\vskip 0.8in plus 0.4in
   \leftline{#1}\supereject\endgroup}

\def\date#1{\finishtitlepage{#1}}

\def\nolabels{\def\eqnlabel##1{}\def\eqlabel##1{}\def\figlabel##1{}%
     \def\reflabel##1{}}
\def\writelabels{\def\eqnlabel##1{%
     {\escapechar=` \hfill\rlap{\hskip.11in\string##1}}}%
     \def\eqlabel##1{{\escapechar=` \rlap{\hskip.11in\string##1}}}%
     \def\figlabel##1{\noexpand\llap{\string\string\string##1\hskip.66in}}%
     \def\reflabel##1{\noexpand\llap{\string\string\string##1\hskip.37in}}}
\nolabels


\global\newcount\secno \global\secno=0
\global\newcount\meqno \global\meqno=1

\def\newsec#1{\global\advance\secno by1
   \xdef\secsym{\the\secno.}
   \global\meqno=1\bigbreak\medskip
   \noindent{\bf\the\secno. #1}\par\nobreak\smallskip\nobreak\noindent}
\xdef\secsym{}

\def\appendix#1#2{\global\meqno=1\xdef\secsym{\hbox{#1.}}\bigbreak\medskip
\noindent{\bf Appendix #1. #2}\par\nobreak\smallskip\nobreak\noindent}


\def\eqnn#1{\xdef #1{(\secsym\the\meqno)}%
     \global\advance\meqno by1\eqnlabel#1}
\def\eqna#1{\xdef #1##1{\hbox{$(\secsym\the\meqno##1)$}}%
     \global\advance\meqno by1\eqnlabel{#1$\{\}$}}
\def\eqn#1#2{\xdef #1{(\secsym\the\meqno)}\global\advance\meqno by1%
     $$#2\eqno#1\eqlabel#1$$}


\def\myfoot#1#2{{\baselineskip=14.4pt plus 0.3pt\footnote{#1}{#2}}}
\global\newcount\ftno \global\ftno=1
\def\foot#1{{\baselineskip=14.4pt plus 0.3pt\footnote{$^{\the\ftno}$}{#1}}%
     \global\advance\ftno by1}


\global\newcount\refno \global\refno=1
\newwrite\rfile

\def\ref{[\the\refno]\nref}
\def\nref#1{\xdef#1{[\the\refno]}\ifnum\refno=1\immediate
 \openout\rfile=refs.tmp\fi\global\advance\refno by1\chardef\wfile=\rfile
 \immediate\write\rfile{\noexpand\item{#1\ }\reflabel{#1}\pctsign}\findarg}
\def\findarg#1#{\begingroup\obeylines\newlinechar=`\^^M\passarg}
 {\obeylines\gdef\passarg#1{\writeline\relax #1^^M\hbox{}^^M}%
 \gdef\writeline#1^^M{\expandafter\toks0\expandafter{\striprelax #1}%
 \edef\next{\the\toks0}\ifx\next\null\let\next=\endgroup\else\ifx\next\empty%
\else\immediate\write\wfile{\the\toks0}\fi\let\next=\writeline\fi\next\relax}}
     {\catcode`\%=12\xdef\pctsign{
\def\striprelax#1{}

\def\semi{;\hfil\break}
\def\addref#1{\immediate\write\rfile{\noexpand\item{}#1}} 

\def\listrefs{\vfill\eject\immediate\closeout\rfile
   \centerline{{\bf References}}\bigskip{\frenchspacing%
   \catcode`\@=11\escapechar=` %
   \input refs.tmp\vfill\eject}\nonfrenchspacing}

\def\startrefs#1{\immediate\openout\rfile=refs.tmp\refno=#1}


\global\newcount\figno \global\figno=1
\newwrite\ffile
\def\fig{\the\figno\nfig}
\def\nfig#1{\xdef#1{\the\figno}\ifnum\figno=1\immediate
     \openout\ffile=figs.tmp\fi\global\advance\figno by1\chardef\wfile=\ffile
     \immediate\write\ffile{\medskip\noexpand\item{Fig.\ #1:\ }%
     \figlabel{#1}\pctsign}\findarg}

\def\listfigs{\vfill\eject\immediate\closeout\ffile{\parindent48pt
     \baselineskip16.8pt\centerline{{\bf Figure Captions}}\medskip
     \escapechar=` \input figs.tmp\vfill\eject}}


\def\letter{\raggedright\parindent=0pt}
\def\endmode{}
\def\longindent{\parindent=3.25truein\obeylines\parskip=0pt}
\def\letterhead{\null\vfil\begingroup
   \parindent=3.25truein\obeylines
   \def\endmode{\medskip\endgroup}}

\def\sendingaddress{\endmode\begingroup
   \parindent=0pt\obeylines\def\endmode{\medskip\endgroup}}

\def\salutation{\endmode\begingroup
   \parindent=0pt\obeylines\def\endmode{\medskip\endgroup}}

\def\body{\endmode\begingroup\parskip=\smallskipamount
   \def\endmode{\medskip\endgroup}}

\def\closing{\endmode\begingroup\longindent
   \def\endmode{\endgroup}}

\def\signed{\endmode\begingroup\longindent\vskip0.8truein
   \def\endmode{\endgroup}}

\def\endofletter{\endmode \ifnum\pageno=1 \nopagenumbers\fi
     \vfil\vfil\eject\end}


\def\noblackbox{\overfullrule=0pt}
\def\inv{^{\raise.18ex\hbox{${\scriptscriptstyle -}$}\kern-.06em 1}}
\def\dup{^{\vphantom{1}}}
\def\Dsl{\,\raise.18ex\hbox{/}\mkern-16.2mu D} 
\def\dsl{\raise.18ex\hbox{/}\kern-.68em\partial}
\def\slash#1{\raise.18ex\hbox{/}\kern-.68em #1}
\def\lspace{}
\def\lbspace{}
\def\boxeqn#1{\vcenter{\vbox{\hrule\hbox{\vrule\kern3.6pt\vbox{\kern3.6pt
     \hbox{${\displaystyle #1}$}\kern3.6pt}\kern3.6pt\vrule}\hrule}}}
\def\mbox#1#2{\vcenter{\hrule \hbox{\vrule height#2.4in
     \kern#1.2in \vrule} \hrule}}  
\def\bar{\overline}
\def\e#1{{\rm e}^{\textstyle#1}}
\def\del{\partial}
\def\curly#1{{\hbox{{$\cal #1$}}}}
\def\curlyD{\hbox{{$\cal D$}}}
\def\curlyL{\hbox{{$\cal L$}}}
\def\vev#1{\langle #1 \rangle}
\def\psibar{\overline\psi}
\def\lform{\hbox{$\sqcup$}\llap{\hbox{$\sqcap$}}}
\def\darr#1{\raise1.8ex\hbox{$\leftrightarrow$}\mkern-19.8mu #1}
\def\half{{\textstyle{1\over2}}} 
\def\roughly#1{\ \lower1.5ex\hbox{$\sim$}\mkern-22.8mu #1\,}
\def\MSbar{$\bar{{\rm MS}}$}
\hyphenation{di-men-sion di-men-sion-al di-men-sion-al-ly}

\def\third{{\textstyle{1\over 3}}}
\def\hf{\half}
\def\nonp{non-perturbative}
\def\hmm{hermitian matrix model}
\def\integ#1#2#3{\int_{#1}^{#2}\!\!\! d#3\ }
\def\O{{\cal O}}
\def\R{{\cal R}}
\def\rline{{\rm I}\!{\rm R}} 
\def\pq{$[P,Q]=1$}
\def\pqq{$[{\tilde P},Q]=Q$ }
\def\nuke{Nucl.Phys.}
\def\pl{Phys.Lett.}
\def\re{\rho_R}
\def\im{\rho_I}
\def\trho{\upsilon}
\def\NP#1{Nucl. Phys.\ {\bf #1}\ }
\def\PL#1{Phys. Lett.\ {\bf #1}\ }
\def\CMP#1{Comm. Math. Phys.\ {\bf #1}\ }
\def\LNC#1{Lett. Nuovo Cimento\ {\bf #1}\ }
\def\NC#1{Nuovo Cimento\ {\bf #1}\ }
\def\CQG#1{Class. Quantum Grav.\ {\bf #1}\ }
\def\PR#1{Phys. Rev\ {\bf #1}\ }
\def\PREP#1{Phys. Rep\ {\bf #1}\ }
\def\PRL#1{Phys. Rev. Lett\ {\bf #1}\ }
\def\JMP#1{J. Math. Phys\ {\bf #1}\ }
\def\SAM#1{Stud. Appl. Math\ {\bf #1}\ }
\def\MPL#1{Mod. Phys. Lett\ {\bf #1}\ }
\def\PA#1{Physica\ {\bf #1}\ }
\def\PP#1{Princeton preprint\ { #1}\ }
\def\RP#1{Rutgers preprint\ { #1}\ }
\def\sec#1{
\bigskip
\noindent{\bf #1}
\bigskip}

\preprint{\vbox{\rightline{SHEP 91/92--25}
\vskip2pt\rightline{G\"oteborg ITP 92--21}}}
\vskip -1.5cm
\title{Global KdV Flows and Stable 2D Quantum Gravity}
\author{Clifford V. Johnson$^1$, Tim R. Morris$^1$
and Anders W\"atterstam$^2$}{}{}
\addressline{\it ${}^1$Physics Department}
\addressline{\it University of Southampton}
\addressline{\it SO9 5NH, U.K.}
\addressline{and}
\addressline{\it ${}^2$Institute for Theoretical Physics}
\addressline{\it Chalmers Institute of Technology}
\addressline{\it S--412 96 G\"oteborg, Sweden}
\vskip -0.5cm

\abstract
The string equation for the  \pqq formulation of non--perturbatively stable
2D quantum gravity coupled to the  $(2m-1,2)$ models is studied. Global KdV
flows between  the appropriate solutions  are considered as  deformations of
two  compatible  linear  problems.  It  is  demonstrated that the necessary
conditions for such flows to exist are satisfied. A numerical study reveals
such flows  between the pole--free  solutions of pure  gravity ($m=2$), the
Lee--Yang edge model ($m=3$) and topological gravity ($m=1$). We conjecture
that this  is the case  for all of  the $m$--critical models.  As the $m=1$
solution is  unique these global flows  define a {\sl unique}  solution for
each $m$--critical model.

\date{April 1992.}

\sec{Introduction}
It is well known\ref\orig{E.Br\'{e}zin and V.Kazakov, Phys.Lett. {\bf B236}
(1990)  144\semi M.Douglas  and S.H.Shenker,  Nucl.Phys. {\bf  B335} (1990)
635\semi D.J.Gross and A.A.Migdal,  Phys.Rev.Lett. {\bf 64} (1990) 127\semi
D.J.Gross   and   A.A.Migdal,   Nucl.   Phys.   {\bf   B}340   (1990)  333.
}\ref\david{F.David, Mod. Phys. Lett. {\bf  A5} (1990) 1019\semi\ F.David,
Nucl.Phys. {\bf B348} (1991) 507.}
that the non--perturbative sector of the
\pq\  formulation of  the $(2m-1,2)$  minimal models  coupled to gravity is
problematic   for  $m$   even.  The   string  equations   for  the   string
susceptibility   $\rho$   of   this   definition   are   the   Painlev\'e~I
hierarchy\orig, $(m+\half)t_m\R_m[\rho]=z$. For $m$ even, the solutions for
$\rho$    with    the    asymptotics    $\rho\rightarrow    z^{1/m}$    for
$z\rightarrow\pm\infty$  are  the  {\sl  complex}  truncated Boutroux--type
solutions\ref\shyamjoe{S.Chaudhuri  and J.Lykken,  \NP{B367}\ (1992) 614.},
which   are  unacceptable.   (Without  loss   of  generality   we  may  set
$t_m=1/(a_m(m+\half))$, where  $a_m$ is the coefficient  of $\rho^m$ in the
Gel'fand--Dikii    polynomial    $\R_m$.)    Real    solutions   with   the
$z\rightarrow+\infty$   asymptotic  above   have  poles,   which  are  also
unacceptable\david.  An   analysis  of  the   Dyson--gas  picture  of   the
matrix--model  realisation of  these critical models\ref\critical{S.Dalley,
\PL{B253}\ (1991)  292\semi S.Dalley, C.V.Johnson  and T.R.Morris, \MPL{A6}
(1991)  439.}  reveals  a  fundamental  instability  in the critical matrix
eigenvalue configurations for $m$ even.

The     models     have      an     underlying     KdV--flow     structure,
$\partial_{t_m}\rho=\R^{'}_{m+1}$, which describes  the evolution of $\rho$
towards   the   $m$th   model   in   the   interpolating   string  equation
$\sum_{m=1}^\infty(m+\half)t_m\R_m=z$. Many attempts  were made to surmount
the problem of the non--perturbative  instabilty of pure gravity ($m=2$) as
defined by the Painlev\'{e}~I  equation. One of these\ref\flow{M.R.Douglas,
N.Seiberg and S.Shenker,  \PL{B244}\ (1990) 381.} was to  try to define the
solution  for the  string susceptibility  by flowing  from the stable $m=3$
solution\ref\bmp{E.Br\'{e}zin,  Marinari  and  G.Parisi,  \pl\  {\bf  B242}
(1990) 35.},  switching on the massive  coupling $t_2$. The result  of this
numerical work showed the flow to  be unstable at large $t_2$: The solution
for the  string susceptibility develops  wild oscillations in  the negative
$z$ region,  signalling the tunnelling  of eigenvalues in  the matrix model
and  presumably heralding  the onset  of the  generic singular  solution of
Painlev\'e~I  in  the  limit.  In  ref.\ref\moore{G.Moore, \CMP{133} (1990)
261.} the theory of isomonodromic deformations was used to study such flows
between solutions  of the string  equations with the  required asymptotics.
This study showed analytically that there  is no path between the $m=2$ and
$m=3$  models  via  the  KdV  flows.  This  is  true  for  all neighbouring
$m$--critical models in this definition\ref\patrik{P.Hermansson, \PL{B263}\
(1991)  385.}.  The  KdV  flows,  although  present  infinitesimally in the
Painlev\'e~I definition, fail to materialize globally.

Later,  another definition  of the  above $(2m-1,2)$  models was  proposed,
which shared the perturbative  results of the previous definition,
but  lacked   the  non--perturbative
instabilities,  as   analysis  of  its  original   matrix  model  definition
shows\ref\multi{S.Dalley,  C.  Johnson  and  T.  Morris,  \NP{B368}  (1992)
625.}\ref\npqg{S.Dalley, C. Johnson and  T. Morris, \NP{B368} (1992) 655.}.
This  is  the   \pqq\  definition, the central  principle  of which is the
non--perturbative preservation of the KdV flows.
This results in the most general
string equations compatible with the KdV flows which are
\eqn\smiley{u\R^2-{1\over   2}\R\R^{''}+{1\over  4}(\R^{'})^2=0}   for  the
string                susceptiblity                $u$,               where
$\R\equiv\sum_{m=1}^\infty(m+\half)t_m\R_m-z$  for  the  interpolating case
and  we  set  $t_m=2/(a_m(2m+1))\delta_{m,0}$  for  the  pure $m$--critical
model.  The  solution  for  $u$  for  such  a  model  has  the  asymptotics
$u\rightarrow   z^{1/m},0$  for   $z\rightarrow\pm\infty$.  These  boundary
conditions can be shown to fix the integration constants so that  the
number of solutions at each  $m$--critical point is discrete\npqg.
Numerical  studies have  revealed\npqg\ref\pqmodels{C.V.Johnson, T.R.Morris
and  B.Spence,  Southampton  preprint  SHEP  90/91--30,  Imperial  preprint
TP/91--92/01}  just one  such solution  for  each  of the  $m=1,2$ and  $3$
models\foot{In the case  of $m=1$, a simple change  of variables\npqg\ maps
to the problem of proving that there exists a  unique pole-free solution of
the Painlev\'e~II
equation. Such a solution  is   known  to  exist\ref\haste{S.P.Hastings  and
J.B.Mcleod,    Arch.   Rat.    Mech.    and    Anal.   {\bf    73}   (1980)
31.}\ref\abloone{M.J. Ablowitz and H.  Segur, Phys.Rev.Lett {\bf 38} (1977)
1103.}.}. In each case the solution was manifestly pole--free.

These results suggest the existence  of a {\sl unique, pole--free} solution
to  each  of  the  $m$--critical  models,  thus  completing  the program of
presenting a definition  of the $(2m-1,2)$ models coupled  to gravity which
is stable  everywhere. In this letter  we wish to address  this suggestion.
There  exist  powerful  techniques\foot{See  for  example  ref.\ref\flas{H.
Flaschka and A. Newell, \CMP{76}  (1980) 65.}. Other references will appear
later in this letter.}\ for the analysis of equations such as \smiley. This
is due to the fact that they are related to non--linear evolution equations
which are solvable via the Inverse Scattering Transform\ref\ist{C.S. Gardner,
J.M. Greene, M.D.
Kruskal and R.M Miura, Comm. Pure Appl. Math. {\bf 27} (1976) 97.},
in  this  case  the  KdV
heirarchy.  Indeed,  as discussed in ref.\ref\unitary{C.V.Johnson,  T.R.Morris
and  A.W\"atterstam,
Southampton preprint SHEP 91/92--19, and G\"oteborg preprint ITP 92--20.}, the
equation \smiley\ for
the $m$th critical model  defines (a sub--class of)  the {\sl self--similar}
solutions of  the $m$th
member of the  KdV heirarchy. Recall that the string equations found  for the
unitary matrix
models\ref\periwal{ V.Periwal  and  D.Shevitz,  Phys.Rev.Lett.   {\bf  64}
(1990)  1326\semi
V.Periwal  and  D.Shevitz,  \NP{B344}  (1990)  731.}  are the self--similar
solutions of  the mKdV hierarchy.  The non--perturbative solutions  for the
family  of  critical   models  for  that  hierarchy  have   been  shown  to
exist\ref\anders{A.W\"{a}tterstam,   \PL{B263}    (1991)   51.}\ref\others{
C.Crnkovi\'c,  M.Douglas  and  G.Moore,  \NP{B360}  (1991)  507.}. This was
carried  out  by  the  use  of  a Gel'fand--Levitan--Marchenko--type linear
integral  equation\ref\glm{V.E.Zakharov  and  A.B.Shabat,  Funct.Anal.Appl.
{\bf  8} (1974)  226\semi I.M.Gel'fand  and B.M.  Levitan, Amer. Math. Soc.
Trans. (1)  {\bf 1} (1953) 253\semi  Z.S.Agranovich and V.A.Marchenko, {\sl
`The Inverse Problem  of Scattering Theory',} New York:  Gordon and Breach,
1963.}\ref\ablotwo{ M.J. Ablowitz, A.  Ramani and H. Segur, \JMP{21} (1980)
715\semi  M.J. Ablowitz,  A. Ramani  and H.  Segur, \JMP{21}  (1980) 1006.}
equivalent to  the string equation. Standard  inverse scattering techniques
are then  used, resulting in the  definition of a one--parameter  family of
pole--free solutions for each model, where  for one value of the parameter,
there is a solution with the required asymptotics.

We  would like  to carry  out a  similar analysis  for our string equations
\smiley. We derive  an integral equation equivalent to  \smiley, but due to
technical difficulties to be described later  it is not clear that using it
we  can construct  arguments which   facilitate a  proof of  uniqueness and
freedom  from poles.  Instead, in  this letter  we take  a different route.
Following the lead of refs.\flas\moore, we formulate our string equation as
a flatness condition for  two first order linear systems.  The KdV flows in
this framework can be shown  to represent isomonodromic deformations of one
of these linear systems and isospectral deformations of the other. We study
the form of the monodromy data  when we impose the asymptotics required for
our  solutions for  the string  susceptiblity. We  find that  there are  no
global obstructions to KdV flows  between the $m$--critical models. Turning
to a numerical study, we demonstrate that such KdV flows are indeed present
by  tracking the  flows between  the $m=1,2$  and $3$  solutions. This then
allows us to  {\sl define a unique solution}  for the string susceptibility
of  each $m$--critical  model by  global  KdV  flow from  the unique  $m=1$
solution.

\def\frac#1#2{{#1\over#2}}
\def\hspace#1{\hskip #1}

\sec{The analysis of the string equation. }
The  starting   point  is  to
rewrite  the  differential  of  \smiley\  as  the  compatibility  condition
$[L,P]=0$  for  two  linear  problems:
\eqn
\lax{\eqalign{  L{\bf  V}  =
0\hspace{1cm}  &  L  =  -\frac{d}{dz}  +  \left(\matrix  {0&\lambda  + u\cr
1&0\cr}\right)\; ,\cr P{\bf V} =  0\hspace{1cm} & P = -\frac{d}{d\lambda} +
\left(\matrix{\alpha   &\beta  \cr\gamma   &-\alpha\cr}\right)\;  }}
where
$\gamma={1\over    {2\lambda}}    (z+\sum_{k=0}^{m}{\lambda}^{k}\R_{m-k})$,
$\alpha=\gamma^{'}/2$, and $\beta=\gamma(\lambda + u ) - \alpha^{'}$
The  compatibility  condition  above  arises  naturally  in  the context of
isomonodromic deformations, reviewed in\ref\jimone{M. Jimbo, T. Miwa and K.
Ueno,  \PA{2D}  (1981)  306.}--\ref\its{A.R.  Its  and  V. Yu. Novokshenov,
Springer Lect.  Notes Math. 1191.}.  We follow the  methods in those  works
closely, and refer the reader to  them for details. The method concentrates
on the second equation in \lax. This equation is singular at the origin and
at infinity and we diagonalize the most singular part at infinity, setting
\eqn
\cov{
{\bf V} = A{\bf W}, \hspace{1cm} A =
\left(\matrix{{\lambda}^{\frac{1}{4}}&-{\lambda}^{\frac{1}{4}}\cr
{\lambda}^{-\frac{1}{4}}&{\lambda}^{-\frac{1}{4}}}\right)
}
We make the change of variable $\zeta^{2} = \lambda$, transforming the
equation into
\eqn
\neweq{
{{d{\bf W}}\over{d\zeta}}=\tilde{M}{\bf W}
}
where now ( $\sigma_{i}$ are the Pauli matrices)
$
\tilde{M} = ({1}/{2\zeta}-2\alpha\zeta )\sigma_{1}
+ i(\beta -\gamma\zeta^{2})\sigma_{2}
 + (\beta +\gamma\zeta^{2})\sigma_{3}.
$
An asymptotic treatment around $\zeta = \infty$
reveals that there exists a formal solution $\Psi$ of \neweq\
\jimone\ref\jimtwo{M. Jimbo and T. Miwa, \PA{2D} (1981) 407.}
\eqn
\formal{
\Psi  =  (1-\frac{u}{4\zeta^{2}}\sigma_{1}+\cdots )\, \e{\left(
\zeta^{2m+1}/2(2m+1)+
x\zeta\right) \sigma_{3}}
}
where the dots  represent higher order terms and  diagonal parts, which are
increasingly (with  $m$) complicated functions of  $u$ and its derivatives.
The  formal  solution  is  such  that  if  we  divide up a neighbourhood of
$\infty$ into sectorial domains $\Omega_{k}$,  then in each domain there is
a true  solution of  \neweq, $\Psi_{k}$,  which is  asymptotic to  \formal.
Furthermore,  on  $\Omega_{k+1}  \cap  \Omega_{k}$,  $\Psi_{k+1} = \Psi_{k}
S_{k}$ for some triangular matrix $S_{k}$ called a {\sl Stokes matrix}. (We
will also use  $S_k$ to label the overlap  $\Omega_{k+1} \cap \Omega_{k}$.)
In   our  case   there  are   $4m+2$  sectors   around  infinity   defined
by ($k=1,2,\ldots  ,4m+2$) $  \Omega_{k}=[\pi /2+{\pi  (k-2)}/(2m+1) <
arg \zeta < \pi /2+{\pi k}/(2m+1)].  $ Each sector $S_{k}$ contains a
unique ray,  ${\pi (k+m)}/{2m+1}$, called  a {\sl Stokes  line} along which
one  of the  fundamental solutions  is maximally  dominant. The initial lines
of the sectors $S_{k}$ are called
{\sl  conjugate  Stokes  lines}, denoted $C_{k}$.
The Stokes  matrices are  generally
triangular\jimone\ and, taking into account the sign of the leading term
$$
S_{2k+1}=\left(\matrix{1&0\cr s_{2k+1}&1\cr}\right),{\hskip 2cm}
S_{2k}=\left(\matrix{1&s_{2k}\cr 0&1 \cr}\right).
$$

As  our equation  is singular  at the  origin we  should in  principle do a
similar analysis  of the solution  in a neighbourhood  of this singularity.
However,  the  original  singularity  before  the  change  of  variables is
regular, so the  analysis is not so complicated. We  return to the original
equation \lax\ and make  the substitution $\lambda={\zeta}^{2}$ and because
the singularity at  the origin is not diagonalizable  we split the solution
matrix $\Phi$  into its components  $\Phi_1^i$ and $\Phi_2^i$,  $i=1,2$ and
study the equation  in terms of these. Keeping only  highest order terms in
${1}/{\zeta}$ we easily find that the fundamental solutions are
$
\Phi_{1,2}=(b(\zeta ),\; a(\zeta )+b(\zeta )ln\zeta)
$
where $a(\zeta)$and  $b(\zeta)$ are holomorphic  in a neighbourhood  of the
origin. The monodromy property of this solution round the origin is then
\eqn\mono{
\Phi(\zeta \e{2\pi i})=\Phi(\zeta )J, \hspace{1cm} J=\left(
\matrix{1 & 2\pi i \cr 0 & 1}\right).}

In order to compare the solutions  at different points we need to introduce
{\sl connection matrices} relating $\Phi$  and $\Psi$. These are defined by
$ \Psi=\Psi^{0}C  $  where  $\Psi^{0}=A^{-1}\Phi$  and  $A$  was defined in
equation~\cov.  The {\sl  monodromy data}  parametrizing the  solutions are
thus the  Stokes matrices $S_{j}$,  the monodromy matrix  round the origin,
$J$,  and the  connection matrix,  $C$.  Not  all of  these parameters  are
independent,  and the  number of  independent parameters  may be reduced by
studying the symmetries  of the matrix $\tilde M$,  and imposing reality on
the string  susceptibility $u$. Further relations  are obtained by studying
the behaviour of the solutions as we  encircle the origin once. Half of the
$4m+2$ original Stokes parameters are  determined by the symmetries and one
by the  behaviour around the origin,  leaving $2m$ parameters. Of  the four
parameters  in the  connection matrix  two are  determined by the behaviour
around  the  origin,  one  by  the  requirement  $detC=1$  and the last one
describes a global scaling freedom in the solution\flas.

In ref.\moore\ the corresponding problem for the Painlev\'e I hierarchy was
studied resulting in necessary conditions for the global KdV flows to exist
between  $m=2$  and  $m=3$  and  these  conditions  were  proven  not to be
satisfied. In short the requirement is that the sectors in which the Stokes
matrices  are  non-trivial  should  overlap  between  the different models,
because   {\sl   the   KdV    flow   equations   preserve   the   monodromy
data}\moore\flas. The  Stokes matrices can  be calculated by  the method in
ref.\its. However  we will not  do so here  as there is  a pictorial way to
proceed in order to derive the necessary conditions. In ref.\moore\ it was
proven  that when  four conjugate  Stokes lines  emerge from  a point,
the Stokes matrix corresponding to the  middle sector must be trivial, i.e.
unity. So we  will impose our boundary conditions and  use this argument to
find the sectors that must be trivial. First we derive necessary conditions
on  the Stokes  matrices so  that $u\sim  z^{1/m}$ as $z\rightarrow\infty$.
Keeping only highest order terms as $z\rightarrow\infty$ we rescale, change
variables  $\zeta=\lambda\tau^{1/2m}$,  $u=\tau^{1/m}$  and  determine  the
components   of  equation~\lax\   as  $\tau\rightarrow\infty$:  $\alpha=0$,
$\beta=\tau(1+\lambda^{2}){1}/{2\lambda^{2}}\sum\lambda^{2k}a_{m-k}$,   and
$\gamma=\tau^{1-{1}/{m}}   {1}/{2\lambda^{2}}\sum\lambda^{2k}a_{m-k}$.  The
$a_m$  are the  coefficients of  $u^m$  in  $\R_m$. As  in the  appendix of
ref.\its\    we    write    the    equation    in    component    form   $$
\frac{d}{d\lambda}(P\frac{d\Psi^{1,2}}{d\lambda})+\tau^{2+\frac{1}{m}}
PQ\Psi^{1,2}=0   $$   where   $P   =   {\tau}/{r(\lambda   )}$   and  $Q  =
(1+\lambda^{2})\sum\sum a_{k}a_{j}\lambda^{2(2m-k-j-1)}$. Here $r(\lambda )$
is also  a polynomial but  its explicit form  is of no  interest to us  and
$\Psi^{1,2}$  are the  vector components  of $\Psi$.  This equation  can be
solved  by  standard  WKB  methods  and  the  approximate  solution  is  $$
\Psi^{1,2}_{(WKB)}=(P^{2}Q)^{-1/4} \exp (\pm \tau^{1+1/2m} \int Q^{1/2} dz)
$$ The important part here, for determining necessary conditions, is simply
the form of the polynomial Q. It  always has a double-zero at the origin, a
simple  zero at  $\lambda=\pm i$  and $2l-2$  distinct zeros  of order  two
elsewhere. We may determine the lay  of conjugate
Stokes lines  by taking into  account their behaviour  at infinity and  the
fact that  they only meet at  zeros of $\int Q^{1/2}  dz$ (they are defined
through  ${\rm Re}  \int Q^{1/2}dz  =  0$).  We can  now use  the pictorial
argument of \moore\ to determine  that the Stokes matrices corresponding to
sectors  $S_{i}$ with  $i$ odd  and $1  < i  < 2m+1$  are trivial\foot{This
result  becomes  a  conjecture  for  $m>3$,  as  in  ref.\moore\ due to the
increasing complexity of the configuration of Stokes lines.}.

If   we   now   consider   solutions   which   decay   rapidly   enough  as
$z\rightarrow-\infty$ the matrix $\tilde{M}$ can be replaced by
$
\tilde{M}  =  (x+{1}/{2}\zeta^{2m})\sigma_{3}
$
and the solutions are simply
$
\Psi^{1,2}=\exp (\pm ({\zeta^{2m+1}}/{2(2m+1)}+z\zeta ))
$.
Using this,  we determine the conjugate  Stokes lines and by  repeating the
argument of ref.\moore\  we find that the Stokes  matrices corresponding to
sectors $S_{i}$ with  $i$ even and $1 <  i < 2m+1$ must be  trivial$^3$. In
this  way  we  find  that  if  the  string  susceptibility $u$ is to behave
asymptotically  as  desired  then  it  is  necessary  that  the  nontrivial
monodromy is  concentrated along the  imaginary axis. The  only non-trivial
Stokes matrices  are $S_{1}, S_{2m+1},  S_{2m+2}$ and $S_{4m+2}$.  
that this is only a necessary condition. For example for the $m$=1 equation
in the  unitary matrix case  the coresponding boundary  conditions actually
completely determine the Stokes parameters\its. In
figs.\fig\stokesi
{The $m=1$ Stokes sectors for
{\bf (a) }$u\sim z$ as $z\rightarrow+\infty$ and
{\bf (b) }$u\sim 0$ as $z\rightarrow-\infty$},
\fig\stokesii
{The $m=2$ Stokes sectors for
{\bf (a) }$u\sim z^{\half}$ as $z\rightarrow+\infty$ and
{\bf (b) }$u\sim 0$ as $z\rightarrow-\infty$}\ and
\fig\stokesiii
{The $m=3$ Stokes sectors for
{\bf (a) }$u\sim z^{\third}$ as $z\rightarrow+\infty$ and
{\bf (b) }$u\sim 0$ as $z\rightarrow-\infty$}\
we display the configurations of conjugate Stokes lines
for the cases $m=1,2$ and $3$.

Thus the necessary condition for the existence of flows between the models,
derived  in ref.\moore,  is satisfied   here. We  will later  confirm their
existence by  the numerical study  of the flows  between the $m=1,2$  and 3
models.

Note that the WKB  solutions above and the structure  of the Stokes sectors
are exactly the same as for  the Painlev\'{e} II hierarchy\anders.
In fact it will be shown in a  later work\unitary\
 that the existence of a {\sl Muira map}
between  the  two  heirarchies  allows  further  study  of the solutions to
equation \smiley.

\sec{The Integral Equation.}

We now consider  briefly the inverse problem to that  above, i.e. given the
Stokes matrices we can compute $u$, deriving an integral equation which can
be studied by the methods of refs.\haste\ablotwo.

We immediately specialise to solutions with the asymptotics that we desire.
Thus,   as   derived   before   we    have   only   the   Stokes   matrices
$S_{1},S_{2m+1},S_{2m+2}$  and  $S_{4m+2}$  (the  sectors  closest  to the
imaginary axis)  non-trivial. An important  observation is that  because of
the  singularity at  the origin  the  monodromy  around this  point is  not
trivial  (as it  is for  the  Painlev\'{e}~II  heirarchy) but  is given  by
equation \mono.  A consequence of this  is a branch cut  along the positive
imaginary    axis.     Therefore    $\Psi^{(i)}_{k}\neq\Psi^{(i)}_{k+4m+2}$
(superscripts   denote   components   and   subscripts   sectors).  Another
consequence is that the relations  among the Stokes parameters which follow
from  the symmetries  of the  equation  are  more complicated  than in  the
Painlev\'{e}  II  case.  The  relations  are  $s_{1}  +s_{4m+2}=s=-2i$  and
$s_{k}=s_{k+2m+1}$ for the Stokes parameters $s_k$.
The  main  idea\foot{See, for example.  ref.\flas\ for details.}\
is  to  use  Cauchy's  theorem  and  relate an integral of
$\Psi^{i}$  along the path
$C_{1}$  to one  along $C_{2}$ (see fig.\fig\conti{The integration
contours used to derive the integral equation})
and to  continue this
around the singular point $\zeta =\infty$. Thus we consider the integral
\eqn\intvi{
\int_{C_{1}}\frac{\Psi^{(1)}_{1}(\xi ,z)\e{\theta}}{\xi -\zeta}d\xi}
%
%
where $\theta =-(\zeta^{2m+1}/2(2m+1)+x\zeta )$ and for simplicity we assume
that  the pole $\zeta$  lies in sector  $S_{1}$. We proceed  by using
Cauchy's theorem
\eqn\intvii{
\eqalign{
\int_{C_{1}}&\frac{\Psi^{(1)}_{1}\e{\theta}}{\xi -\zeta}d\xi =\cr
& 2\pi i\Psi^{(1)}_{1}(\zeta )\e{\theta }-\frac{\pi i}{2m+1}\left(
\matrix{1\cr 0\cr}\right)+
\int_{C_{2}}\frac{\Psi^{(1)}_{1}\e{\theta}}{\xi -\zeta}d\xi +
\int_{\gamma_{1}}\frac{\Psi^{(1)}_{1}\e{\theta}}{\xi -\zeta}d\xi.}}
The first term on  the right hand side is the residue  at $\xi =\zeta$, the
second term arises  from an integral at infinity and
$\gamma_{1}$ is a segment
of a circle around the origin.
Next we use the relations among the solutions in neighbouring sectors
$\Psi^{(1)}_{1}=
\Psi^{(1)}_{2}-s_{1}\Psi^{(2)}_{2}$ to rewrite the third term in
equation \intvii
$$\int_{C_{2}}\frac{\Psi^{(1)}_{1}\e{\theta}}{\xi -\zeta}d\xi =
\int_{C_{2}}\frac{\Psi^{(1)}_{2}-s_{1}\Psi^{(2)}_{2}}
{\xi -\zeta}\e{\theta}d\xi\; .$$
Note that $\Psi^{(2)}_{2}\e{\theta}\rightarrow\infty$ as $\zeta
\rightarrow\infty$ in $S_{2}$ and therefore we can not relate the
term $-s_{1}\int_{C_{2}}\frac{\Psi^{(2)}_{2}\e{\theta}}{\zeta -\xi}
d\xi$ by contour integration to an integral along $C_{3}$.
%
%
This integral appears in this form in the final equation.
We can however express the other integral along $C_{2}$
in terms of an integral along $C_{3}$
\eqn\intx{
\int_{C_{2}}\frac{\Psi^{(1)}_{2}\e{\theta}}{\xi -\zeta}d\xi =
-\frac{\pi i}{3}\left(\matrix{1\cr 0\cr}\right)+
\int_{C_{3}}\frac{\Psi^{(1)}_{2}\e{\theta}}{\zeta -\xi}d\xi +
\int_{\gamma_{3}}\frac{\Psi^{(1)}_{2}\e{\theta}}{\zeta -\xi}d\xi}
where now $\Psi^{(1)}_{2}=\Psi^{(1)}_{3}$.
We then continue in this manner around infinity and sum
all the resulting equations to find
\eqn\intxii{
\eqalign{
\Psi^{(1)}(\zeta )&\e{\theta (\zeta )}  =  \cr
& \left(\matrix{1\cr 0}
\right)
-\frac{1}{2\pi i}\int_{C}\frac{\Psi^{(1)}\e{\theta}}{\zeta -\xi}
+\frac{s_{1}}{2\pi i}\int_{C_{a}}\frac{\Psi^{(2)}\e{\theta}}{\zeta -\xi}
d\xi
-\frac{s}{2\pi i}\int_{C_{b}}\frac{\Psi^{(2)}\e{\theta}}{\zeta
-\xi} d\xi}}
where $\Psi^{(i)}=\Psi^{(i)}_{1}$.
The main difference between this equation and the corresponding equation in
case of the Painlev\'{e}~II hierarchy is that the integrals along $C_{1}$
and  $C_{4m+3}$ (on  the two  sides of  the branch  cut along the imaginary
axis)  no longer  cancel. The  contours $C_{a}$  ($C_{b}$) run inward along
$C_{2m+2}$ then  clockwise (anticlockwise) around  the origin and  out
along  $C_{2}$ ($C_{4m+3})$.  The contour  $C$ starts  at $\zeta =\infty$
travels to the left of the branch  cut, encircles the origin and returns to
infinity  along the  other side  of the  branch cut.
A corresponding equation
can be derived for the other component $\Psi^{(2)}$
\eqn\intxiv{\eqalign{
\Psi^{(2)}(\zeta )&\e{-\theta (\zeta )}  =  \cr
& \left(\matrix{0\cr 1}
\right)
-\frac{1}{2\pi i}\int_{C}\frac{\Psi^{(2)}\e{-\theta}}{\zeta -\xi}
 -\frac{s_{1}}{2\pi i}\int_{C_{c}}\frac{\Psi^{(1)}\e{-\theta}}{\zeta -\xi}
d\xi
-\frac{ss_{1}}{2\pi i}\int_{C_{b}}\frac{\Psi^{(2)}\e{-\theta}}{\zeta
-\xi} d\xi}}
where $C_{c}$ runs inward along $C_{2m+3}$ and out along $C_{4m+3}$.
Thus  equations  \intxii\  and  \intxiv\  are  the integral equations which
correspond to  our string equation  for the solutions  with the asymptotics
$u(z)\sim  z^{1/m},0$  as   $z\rightarrow\pm\infty$.  (By  comparison  with
ref.\flas\ we find that these equations
are  precisely   the  same  as   in  the  case   of  the  Painlev\'{e}~II
heirarchy  with a  non-zero constant  on  the  right hand  side. This  is a
further consequence  of the existence of  a map between that  heirarchy and
the  string equations  \smiley. The  significance of  these results will be
discussed in ref.\unitary.)
  In principle  these integral
equations can now be analysed using the methods in \flas\haste\abloone. Due
to  the  complications  from  the  singularity  at  the  origin, we find it
difficult  to  proceed  via  this  route.  To  study  the uniqueness of the
required solutions, we turn once again to the study of the KdV flows.

\sec{The existence of global KdV flows.}
The flow between  solutions of two critical points $m$  and $n$ is governed
by  the  massive  interpolating  string  equation  which  is  \smiley\ with
$\R=(m+\half)t_m\R_m+(n+\half)t_n\R_n-z$. In flowing from  $m$ to $n$, $t_m$
is
redundant and we may rescale variables to set it to $1/(a_m(m+\half))$, where
$a_m$ is the coefficient of  $u^m$ in $\R_m$. $t=(n+\half)a_nt_n$ defines the
one--parameter flow $\partial_tu\sim\R^{'}_{n+1}$. Starting at the solution
$u_m(z)$ for  the pure $m$--critical  theory we evolve  $t$ from zero,  and
track the solution  $u(z,t)$ to the interpolating equation,  using the fact
that the solution can only change infinitessimally via tha KdV flows\npqg.
 Scaling this
solution  to  ${\tilde  u}({\tilde  z},t)=t^{-2/2n+1}u(zt^{-1/2n+1},t)$, we
arrive  at  the  solution  to  the  pure  $n$--critical  model in the limit
$t\rightarrow+\infty$:  $u_n({\tilde  z})=\lim_{t\rightarrow+\infty}{\tilde
u}({\tilde z},t)$.

This  is  precisely  the  procedure  that  we  carried  out.  We set up the
interpolating string  equation as a  two--point boundary value  problem for
arbitrary $t$. We  solved the first differential of  the string equation as
this allows for better numerical behaviour. Otherwise it is necessary to
calculate  $1/\R$  to  evaluate  the  highest  derivative  in \smiley,
which results in
rounding errors for large positive $z$. We ensured that we
obtained solutions  to the correct  equation \smiley\ by  comparison with the
pure $m$--critical  solutions obtained in previous  numerical treatments of
this  string   equation\foot{The  pure  $m=2$  solution   is  displayed  in
ref.\npqg, and  the pure $m=3$  in ref.\pqmodels.}. Here,  we calculate the
positive  $z$   boundary  conditions  at  arbitrary   $t$  by  solving  the
tree--level  equation $u^m+tu^n=z$  numerically. The  negative $z$ boundary
condition was is $u=0$ throughout.

We set  up the flow from  $m=3$ to $m=2$ and  from $m=2$ to $m=1$.  At each
$t$, the  differential equation was  solved after typically  12 iterations,
using the  NAG {\sl FORTRAN}  library routine D02RAF,  with a global  error
estimate of  $\sim 10^{-7}$ in the  case of $m=2\rightarrow m=1$  and $\sim
10^{-4}$  for $m=3\rightarrow  m=2$. Some  of the  intermediate stages  are
displayed  in figs.\fig\flowone{Some  snapshots of  the evolution  from the
$m=2$ (pure  gravity) model (top curve)  to the $m=1$ model.  The values of
the KdV--flow parameter are 0.0, 0.5, 1.0, 2.0, 5.0, 10.0, 30.0, 50.0, 70.0
and 150.0. A rescaling of the final  curve, as described in the text, shows
rapid  convergence  to  the  pure  $m=1$  solution.}\ and \fig\flowtwo{Some
snapshots of the  evolution from the $m=3$ (critical  Lee--Yang) model (top
curve) to the  $m=2$ model. The values of the  KdV--flow parameter are 0.0,
0.5, 1.0,  2.0, 5.0, 10.0, 30.0,  50.0, 70.0 and 150.0.  A rescaling of the
final curve, as described in the  text, shows rapid convergence to the pure
$m=2$ solution.}, starting at $t=0$ and  going up to $t=150$. For large $t$
($>$100) the  routine required an increasingly  higher number of iterations
to meet its error tolerances.
However upon comparison of ${\tilde u}({\tilde z},t)$ as defined above with
the pure solutions it was observed that the limit procedure described above
was  converging rapidly;  well before  any numerical  problems due  to very
large $t$ could develop.

The above convergence of the procedure  is in striking contrast to the flow
in ref.\flow\  using the $\R_m=z$  definition. There the  instabilty of the
flow at large $t$ showed that the KdV flows could not connect the solutions
in the limit. Here, we  have successfully demonstrated the opposite: Global
KdV flows exist  between the pole--free
$m=1,2$ and 3 models.  We find it a reasonable
conjecture that this is true for all the $m$--critical models.

The existence of these global flows  defines for us a {\sl unique} solution
to each of the $m$--critical models by flow from the unique $m=1$ solution.

\bigskip \bigskip \noindent {\bf Acknowledgements}

\noindent
C.J. thanks the S.E.R.C. for
financial support.

\listrefs
\listfigs
\bye